\documentclass[a4paper]{mem}
\usepackage{natbib}
\usepackage{graphicx}
\begin{document}
 \title{Connection between the X-ray, UV and optical emission line
regions of AGNs
}

   \author{L. \v C. Popovi\'c \inst{1,2}%\fnmsep
}

   \offprints{L. \v C. Popovi\'c}
\mail{lpopovic@aob.bg.ac.yu}

   \institute{Astronomical Observatory, Volgina 7, 11160 Belgrade, Serbia;
\email{lpopovic@aob.bg.ac.yu}\\ 
               \and Astrophysicalisches Institut Potsdam, An der
Sternwarte, 14482 Potsdam, Germany; \email{lpopovic@aip.de}
 }

   \abstract{
    In order to  investigate the connection between the X-ray, UV
and optical
spectral line regions, we selected a sample of AGN which have the strong
Fe
K$\alpha$ line and
analyze the UV and optical spectral  lines.
 Here we present  an example of
  AGN NGC 3516. In this object it seems that geometries of all three
emitting regions are similar, i.e. 
 the broad line shapes can be explained by the multicomponent emitting
line
region with a disk
emission. The  disk emission contributes to the wings, while one
(spherical) emitting region contributes  to the line
core. Concerning the Fe K$\alpha$, UV and Balmer line
shapes it seems that these broad line emitting regions in NGC 3516 are
geometrically similar, i.e. in all
wavelengths the lines are emitted partly  from the disk partly from a
spherical region.
   \keywords{Active galactic nuclei --
                spectroscopy -- spectral lines --
                accretion disk
               }
   }
   \authorrunning{L. \v C. Popovi\'c}
   \titlerunning{Connection between the X-ray, UV/optical emission}
   \maketitle
%
%________________________________________________________________

\section{Introduction}

 According to the standard model of Active Galactic Nuclei (AGNs), an AGN
consists of a black hole
surrounded by a
 (X-ray and optical) continuum emitting region probably with an accretion
disk geometry, 
 the Broad Line Region (BLR) and a larger region that usually is referred
to as the Narrow Line Region
(NLR). Variability studies 
 of QSOs indicate  that the size of the X-ray emission region is the order
of
10$^{14-15}$ cm (e.g. Oshima et al. 2001). An Fe K$\alpha$ fluorescence
 line detected in several AGNs near 6.4 keV is thought to originate from
 within a few 10s of   gravitational radii (Fabian et al. 1995). This
line is
thought  to be 
a  fluorescence line of Fe due to emission from a cold or ionized
 accretion disk that is illuminated from a source of hard X-rays
 originating near the central object. 

%                                     Two column figure (place early!)
%______________________________________________ Gamma_1 (lg rho, lg e)
   \begin{figure*}
   \centering
   \resizebox{\hsize}{!}{\rotatebox[]{0}{\includegraphics{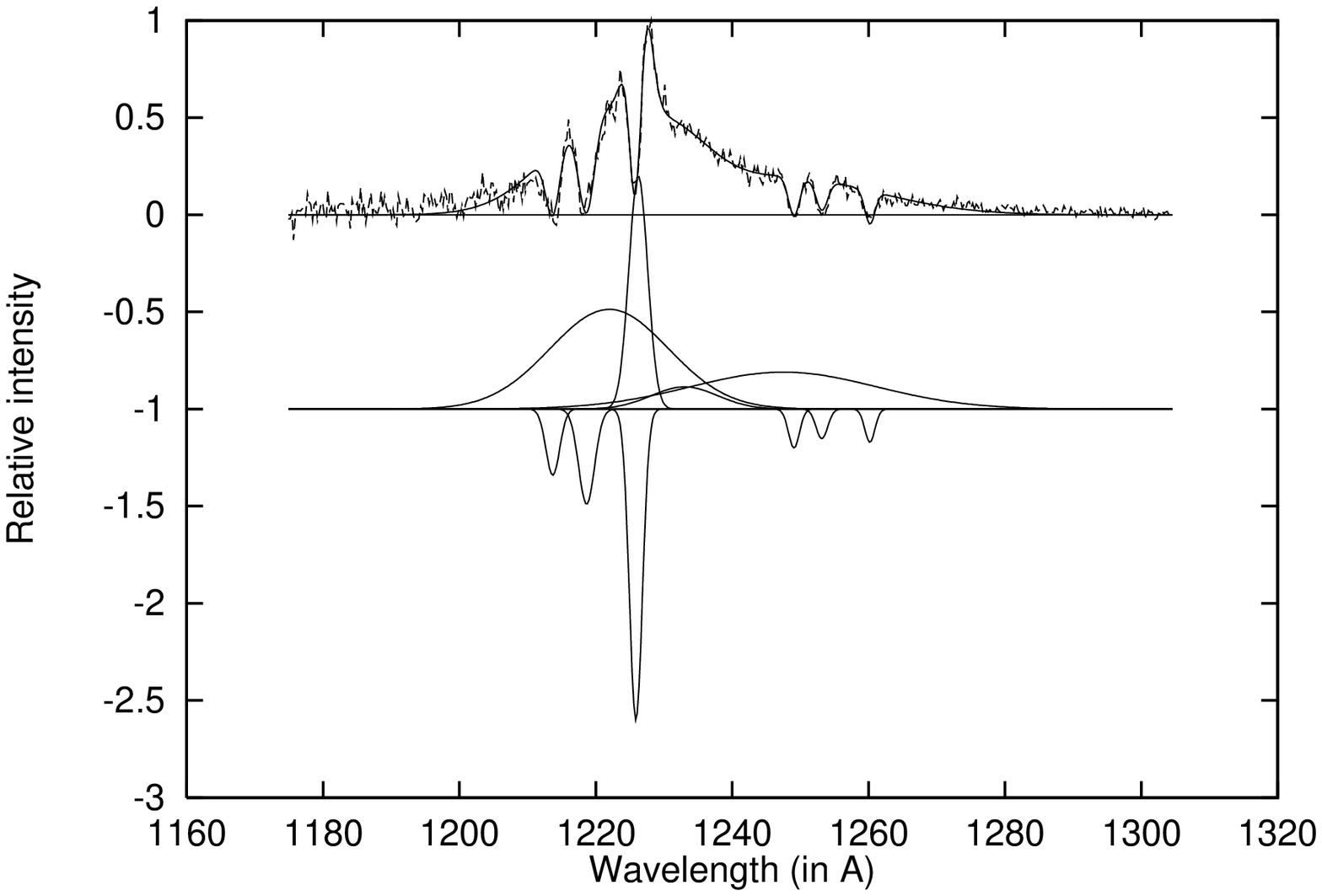}
   \includegraphics{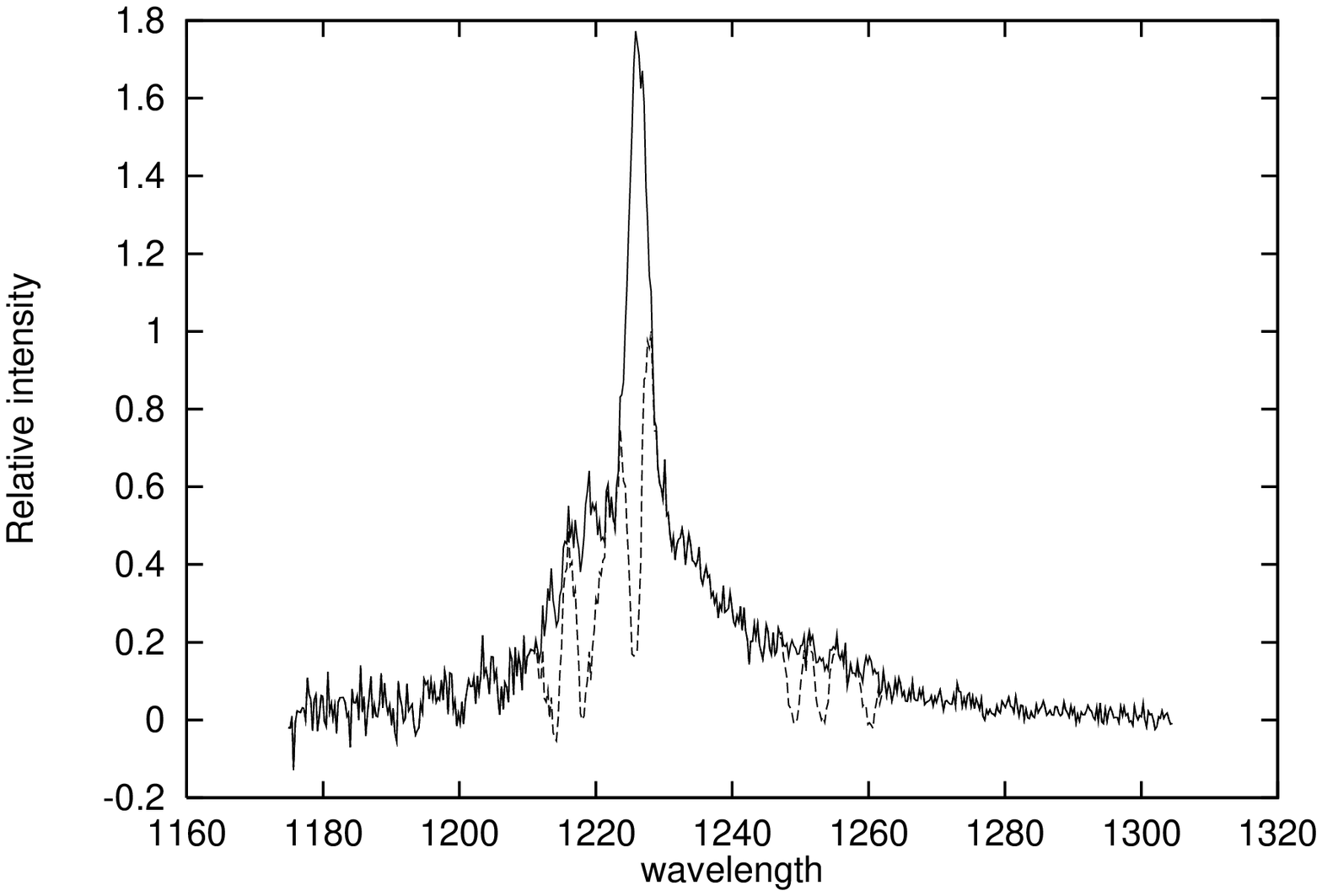}}}
   \caption{{it Left}: The best fit of the Ly$\alpha$ line - the 
Gaussian
components  are presented at bottom.
 {\it Right}: The Ly$\alpha$ line after the absorption
components were subtracted. }
              \label{FigGam}%
    \end{figure*}

 On the other side, the BLR has  also complex physics and kinematics. 
 Kinematics in BLR is widely discussed (see Sulentic et
 al. 2000, and references therein); assuming spherical, cylindrical and
disk
 geometry (see e.g. Popovi\'c et al. 2004, and references
 therein). Although the unified AGN 
 model  (e.g. Elvis 2000) seems to be appropriate to  
 explain  the nature of 
 AGNs, the connection between different line
emission
 regions is not yet clear, e.g.  
 similarity between  geometries of the  X-ray line emitting region and
 the BLR.
Possible  relationships between the emission regions observed in
different  wavelengths,	 were investigated in several works in the
past: it  was found that the steeper slopes of X-ray continua in AGNs are
related to  the presence of narrower optical lines (Boller et
al. 1996).  Furthermore, a correlation was found between the H$\beta$ line
width
(narrower) and the ratio (higher) of the NV/CIV UV lines (Wills et al. 1999). 

Here we present our analysis of the broad UV line shapes of  NGC 3516.
The aim of this work is to find any evidence of  similar geometry between
different emission line regions (X-ray, UV and optical).

%----------------------------------------------------------- S_vib
  \begin{figure}
   \centering
   \resizebox{\hsize}{!}{\includegraphics{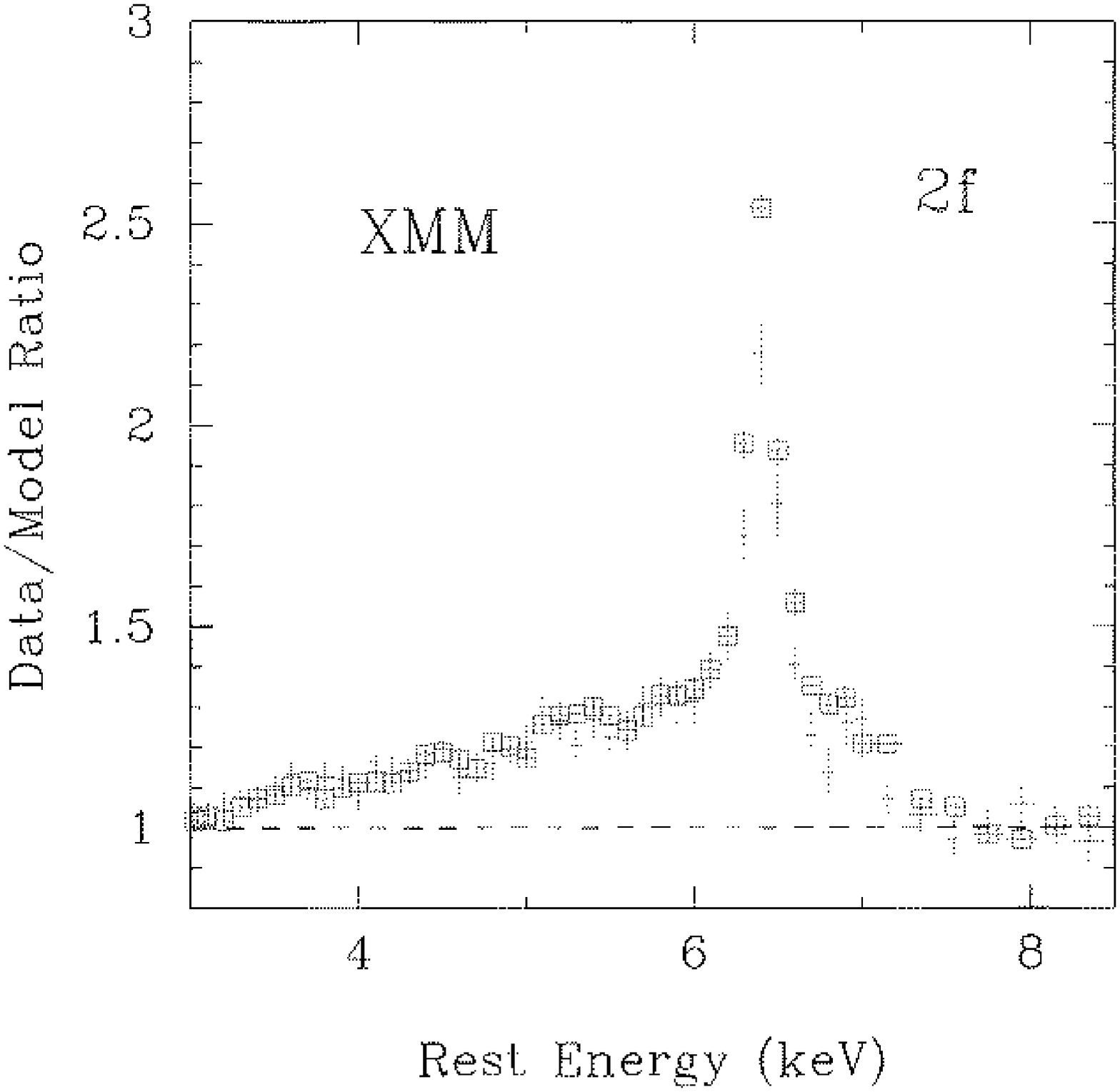}}
   \caption{The observed Fe K$\alpha$ (squares) fitted with the
multi  
component model (crosses)
given by Turner et al. (2002)}
     \label{fit4}
\end{figure}
%
%______________________________________________________________   

\section{The broad lines of NGC 3516}

The existence of an accretion disc in NGC 3516 is  supported  by the
shape of the Fe K$\alpha$ line (Nandra et al. 1997, 1999, Turner et
al. 2002).
The Fe K$\alpha$ line has been modeled according to a disc
geometry. Nandra et al. (1997, 1999) fitted an integrated Fe K$\alpha$ 
profile
deriving the disc inclination for the Schwarzschild metric 35$^{+1}_{-2}$
degrees and for the
Kerr metric 0$^{+19}_{-0}$ degrees. Also, Pariev et al. (2001)
 studied the Fe K$\alpha$ emission obtaining a disc inclination
of about 27 degrees. More recently, Turner et al. (2002) found that the Fe
K$\alpha$ shape has several peaks (5.6, 6.2, 6.4, 6.5, and 6.8) and
alternatively they explained that
peak at 5.6 keV could be the
red horn of a disk line at 
   35R$_g$ (where a weak 6.8 keV peak may be the blue horn associated with
this). Also, they supposed that the 6.2 and 6.5 keV peaks could be due to
emission at 175R$_g$, and that these may be outside the disk
structure. This model can well fit the Fe K$\alpha$ line (Fig. 2).
 
On the
other hand, the H$\alpha$ line was fitted in an earlier work by Sulentic
et al. (1998) with a model including a disc with an inclination about 20
degrees. Finally, Popovi\'c et al. (2002) found that all Balmer lines in
NGC 3516 have disk-like wings. They fit all Balmer lines using the
two-component model and found that the inclination of the optical disk
is around $11\pm 5$ degrees which is comparable with the results of Nandra
(1999)  and Sulentic (1998). Concerning these investigation the Balmer
emission line disk is located from $\sim$ 400 R$_g$ to  $\sim$ 1500
R$_g$. 

Here we present our investigations
of the  broad Ly$\alpha$ line.

\subsection{The UV lines; observation and analysis}

We use HST observations obtained with the FOS and STIS/FUV-MAMA,  covering
the
wavelength
ranges 1150-1730 \AA . The observations were made in 1996 (FOS) and on Apr
13-18
1998 (STIS/FUV-MAMA), in the frame of the campaign monitoring of
NGC 3516 with HST, RXTE and ASCA, for more details see Edelson et
al. (2000). The spectra
were reduced by the HST team. We transform  the wavelength
scale to zero redshifted taking into account the cosmological red-shift
(z=0.00884).
Absorption components are very often present in the UV lines. In order to 
subtract the absorption components, a multi-Gaussian analysis was used
(see
Fig. 1, left). Subtracting the absorption component we reconstruct the
line profile (see e.g. for the Ly$\alpha$, Fig. 1, right).

After that, the lines were fitted by two-component model described in
Popovi\'c et al. (2002,2003,2004) assuming that the emissivity index
$p=3$.

%______________________________________________________________
%----------------------------------------------------------- S_vib
  \begin{figure}
   \centering
   \resizebox{\hsize}{!}{\includegraphics{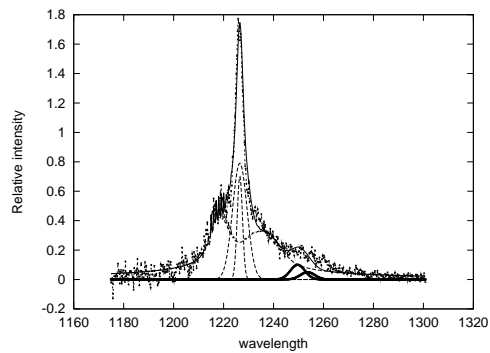}}
  \caption{The observed  Ly$\alpha$ fitted with the two-component
model. The central narrow component probably comes from NLR. The NV lines
are present in the red wing. } 
     \label{fit5}
\end{figure}
%
%______________________________________________________________
%----------------------------------------------------------- S_vib
  \begin{figure}
   \centering   
   \resizebox{\hsize}{!}{
   \includegraphics{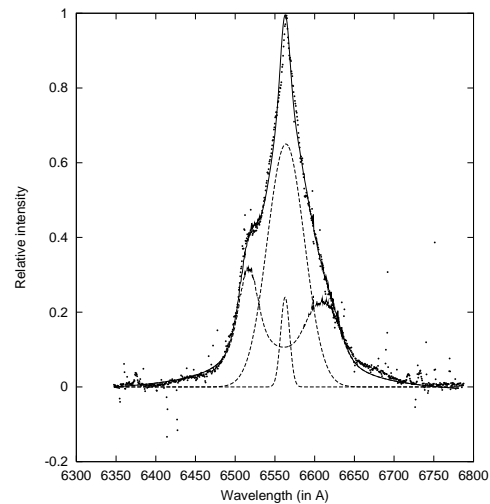}}
  \caption{The observed H$\alpha$ line (dots) fitted with multicomponent
model (solid line) given by  Popovi\'c et al. (2002). With dashed lines
the disk, broad and narrow spherical components are presented.}
     \label{fit6}
\end{figure}
%
%______________________________________________________________   

\section{Results and discussion}

The two-component model  can very well fit the
observed broad UV line profiles, but it is very hard to obtain the disk
parameters
without imposing at
least one constraint because of the large number of parameters and the
lack of two peaks in the line profiles. They can be only roughly estimated
using fitting tests, but when we fixed $p=3$  we found that the 
inclination is
similar to the case of Balmer emission lines (Popovi\'c et
al. 2002) $\approx$ 10$^\circ$, and that the part of emission disk which
emits in the UV lines seems to be
located from $\sim$ 100 R$_g$ to $\sim$ 900 R$_g$. As one can see in
Fig. 3, the Ly$\alpha$ line can be well fitted with multi-component model,
where one component, contributes to the wings, coming from disk.
 Moreover,  the Fe K$\alpha$
and Ly$\alpha$ wings are quit similar (see Fig. 2 and 3), they have an
extended red side (that
in the case of the Ly$\alpha$ cannot be explained by contribution of NV
lines) in both of the lines. In Table 1 we give
the estimated disk parameters for the Fe K$\alpha$, Ly$\alpha$ and Balmer 
lines.

\begin{table}
\begin{center}
\caption[]{The disk parameters for the Fe K$\alpha$, Ly$\alpha$ and Balmer
lines. Reference: (1)- Turner et al. 2002, (2)s,(2)k - Nandra et al. 1999
for Schwarzschild and Kerr metric, respectively, (3) -
Popovi\'c et al. 2002. The inner (R$_{inn}$) and outer (R$_{out}$) disk
radii are given in R$_g=GM/c^2$.}
 \begin{tabular}{|c|c|c|c|c|c|}
\hline
Line(s) & R$_{inn}$  & R$_{out}$  & $i$& Ref. \\
\hline \hline
Fe K$\alpha$ & 3-5 & 35 (175) & 38 & (1) \\
Fe K$\alpha$  & 6 & 80 & 35 & (2)s \\
Fe K$\alpha$  & 2.8 & 400 & 0-19 & (2)k \\
Ly$\alpha$ & $\sim$ 100& 900& $\sim$ 10& \\
Balmer& 400 & 1500& 11& (3) \\
\hline
\end{tabular}
\end{center}
\label{tabl1}
\end{table}

As one can see from the Table 1. the Fe K$\alpha$ disk region should be
more compact than the UV/optical ones. On the other side, the inclination
tends to be small ($i<38^\circ$) and it is in good agreement between
the estimates given by Nandra
et al. (1999) for the  Kerr metric and our estimates, although one should
not
exclude the existence of a warped disk.

\section{Conclusion}

Here we present the investigation of the geometry of different emission
line
regions (Fe K$\alpha$, UV and optical) of NGC 3516. We found that the
broad Ly$\alpha$
line can be well fitted by the two-component model. From this fit we found
the
parameters of the UV disk. These results, together with previous for the
Fe K$\alpha$ (Nandra
et al. 1999; Turner et al. 2002, see Fig. 2) and Balmer lines
(Popovi\'c et al. 2002, see Fig. 4) indicate that geometries of the broad 
X, UV
and optical
emission line regions in NGC 3516 are similar; i.e. that there is a disk
component which contributes to the line wings and one component which
contributes to the line core (see Figs. 2-4).

\begin{acknowledgements}
      This work is a part of the project P1196 supported by the
Ministry of Science and Environmental Protection
of Serbia. The author is supported by the Alexander von Humboldt
foundation through the program for foreign scholars. 

\end{acknowledgements}

\bibliographystyle{aa}

\begin{thebibliography}{}

\bibitem[{Boller et al. (1996)}]{boller} 
Boller, T., Brandt, W. N., Fink, H. 1996, A\&A 305, 53
\bibitem[{Edelson et al. (2000)}]{edel}
Edelson, R., Koratkar, A., Nandra, K., et al. 2000, ApJ, 534, 180
\bibitem[{Elvis (2000)}]{elvis}
Elvis, M. 2000, ApJ, 545, 63
\bibitem[{Fabian et al. (1995)}]{fabian} 
Fabian, A.~C., Nandra, K., Reynolds, C.~S., et al., 1995, MNRAS, 277L, 11
\bibitem[{Nandra et al. (1997)}]{nan1}
Nandra, K., George, I. M., Mushotzky, R. F., Turner, T. J., 
Yaqoob, T. 1997, ApJ, 477, 602
\bibitem[{Nandra et al. (1999)}]{nan2} 
Nandra, K., George, I. M., Mushotzky, R. F., Turner, T. J., Yaqoob,
T. 1999, ApJ, 523, 17
\bibitem[{Oshima et al. (2001)}]{oshima} 
Oshima, T., Mitsuda, K., Ota, N., Yonehara, A., Hattori, M., Mihara,  
T., Sekimoto, Y. 2001, ApJ, 551, 929
\bibitem[{Popovi\'c et al. (2002)}]{pop1}
Popovi\'c, L. \v C., Mediavlilla, E. G., Kubi\v cela, A.,  Jovanovi\'c,
P. 2002, A\&A, 390, 473
\bibitem[{Popovi\'c et al. (2003)}]{pop2}
Popovi\'c, L. \v C., Mediavlilla, E. G., Bon, E., Stani\'c, N.,  Kubi\v
cela, A. 2003, ApJ, 599, 185
\bibitem[{Popovi\'c et al. (2003)}]{pop3}
Popovi\'c, L. \v C., Mediavlilla, E. G., Bon, E., Ili\'c, D. 2004,  A\&A,
423, 909
\bibitem[{Sulentic et al. (1998)}]{sule1}          
Sulentic, J. W., Marziani, P., Zwitter, T., Calvani, M., 
Dultzin-Hacyan, D. 1998, ApJ, 501, 54
\bibitem[{Sulentic et al. (2000)}]{sule2}
Sulentic, J. W., Marziani, P.,  Dultzin-Hacyan, D. 2000, ARA\&A, 38, 521 
\bibitem[{Turner et al. (2002)}]{tur}
Turner, T. J., Mushotzky, R. F., Yaqoob, T., et al. 2002, ApJ,
574, 123
\bibitem[{Wills et al. (1999)}]{wills}
Wills, B.J., Laor, A., Brotherton, M. S., et al. 1999, ApJ, 515, 53

\end{thebibliography}

\end{document}